\documentclass[12pt]{article}

\advance \topmargin by -\headheight
\advance \topmargin by -\headsep

\evensidemargin \oddsidemargin
\begin{document}

\begin{titlepage}
\vspace*{-1cm}

\vskip 3cm

\vspace{.2in}
\begin{center}
{\large\bf  The Beauty of Self-Duality\footnote{Published in {\em Tribute to Ruben Aldrovandi}, editors F. Caruso, J. G. Pereira and A. Santoro; pages 219-240;  Editora Livraria da F\'isica, S\~ao Paulo (2024)}}
\end{center}

\vspace{.5cm}

\begin{center}
L. A. Ferreira\footnote{laf@ifsc.usp.br}

\vspace{.3 in}
\small

\par \vskip .2in \noindent
Instituto de F\'\i sica de S\~ao Carlos; IFSC/USP;\\
Universidade de S\~ao Paulo, USP  \\ 
Caixa Postal 369, CEP 13560-970, S\~ao Carlos-SP, Brazil\\

\vskip 2cm

{\em Dedicated to the Memory of Prof. Ruben Aldrovandi}

\vskip 2cm

\begin{abstract}

Self-duality plays a very important role in many applications in field theories possessing topological solitons. In general, the self-duality equations are first order partial differential equations such that their solutions satisfy the second order Euler-Lagrange equations of the theory. The fact that one has to perform one integration less to construct self-dual solitons, as compared to the usual topological solitons, is not linked to the use of any dynamically conserved quantity. It is important that the topological charge admits an integral representation, and so there exists a density of topological charge. The homotopic invariance of it leads to local identities, in the form of second order differential equations. The magic is that such identities become the Euler-Lagrange equations of the theory when the self-duality equations are imposed. We review some important structures underlying the concept of self-duality, and show how it can be applied to kinks, lumps, monopoles, Skyrmions and instantons.

\end{abstract}

\normalsize
\end{center}
\end{titlepage}

\section{Introduction}
\label{sec:introduction}
\setcounter{equation}{0}

Topological solitons play a fundamental role in the study of non-linear phenomena in many areas of science. The  non-trivial topological structures make them quite stable, and consequently very important in the description of many facets of the theory. Topological solitons appear in  a variety of theories ranging from kinks in $(1+1)$-dimensions, to vortices  in $(2+1)$-dimensions, magnetic monopoles and Skyrmions in $(3+1)$-dimensions, and instantons in four dimensional Euclidean spaces. They are  relevant for many non-linear phenomena in high energy physics, condensed matter physics and science in general \cite{mantonbook,shnirbook,shnirbookmonopoles}. 

Among the types of topological solitons there is a class which is special, the so-called self-dual solitons. They are classical solutions of the self-duality equations which are first order differential equations that imply the second order Euler-Lagrange equations of the theory. In addition, on each topological sector  there is a lower bound on the static energy, or Euclidean action, and the self-dual solitons saturate that bound. Therefore, self-dual solitons are very stable.

The reason why one  performs just one integration to construct self-dual solitons, instead of two in the case of the usual topological solitons, is  not linked to  dynamically conserved quantities. In all cases where self-duality is known to work, the relevant topological charge admits an integral representation, and so there exists a density of topological charge. As such charge is invariant under any smooth (homotopic) variations of the fields, it leads to local identities, in the form of second order differential equations, that are satisfied by any regular configuration of the fields, not necessarily solutions of the theory. The magic is that such identities become the Euler-Lagrange equations of the theory when the (first order) self-duality equations are imposed. 

The concept of generalized self-dualities has been put forward using such an ideas where one can construct, from one single topological charge, a large class of field theories possessing self-dual sectors \cite{genbps}. In $(1+1)$-dimensions it was possible to construct field theories, with any number of scalar fields, possessing self-dual solitons, and so generalizing what is well known in theories with one single scalar field, like sine-Gordon and $\lambda\, \phi^4$ models \cite{sd2d,bpscomments}. In addition, exact self-dual sectors were constructed for Skyrme type theories by the addition of extra scalar fields \cite{bpswojtek,bpsshnir,laf2017,us}, and concrete applications have been made to nuclear matter \cite{nuclear}. 

In this paper we review those developments in a simple and concise way. The concept of self-duality has been used for a long time in several contexts \cite{bogo,prasad,belavincp1,belavininstanton}, and we give here the main idea behind the concept of generalized self-duality proposed in \cite{genbps}, and in fact genereralizing it to the case of complex fields.  Consider a field theory that possesses a topological charge with an integral representation of the form
\begin{equation}
Q=\frac{1}{2}\,\int d^dx\,\left[ {\cal A}_{\alpha}\,{\widetilde{\cal A}}_{\alpha}^*+{\cal A}_{\alpha}^*\,{\widetilde{\cal A}}_{\alpha}\right]
\label{topcharge}
\end{equation}
where  ${\cal A}_{\alpha}$ and ${\widetilde{\cal A}}_{\alpha}$ are functionals of the fields of the theory and their first derivatives only, and where $^*$ means complex conjugation, and not transpose complex conjugate.  The index $\alpha$ stands for any type of indices, like vector, spinor, internal, etc, or groups of them. The fact that $Q$ is topological means that it is invariant under any smooth (homotopic) variations of the fields. Let us denote the fields by $\chi_{\kappa}$, and they can be scalar, vector, spinor fields, and the index $\kappa$ stands for the space-time and internal indices. We take $\chi_{\kappa}$ to be real, and so, if there are complex fields, $\chi_{\kappa}$ stands for the real and imaginary parts of those fields. The invariance of $Q$ under smooth variations of the fields, i.e. $\delta\,Q=0$, leads to the identities
\begin{eqnarray}
&&
\frac{\delta\, {\cal A}_{\alpha}}{\delta\,\chi_{\kappa}}\, {\widetilde{\cal A}}_{\alpha}^*-\partial_{\mu}\left(\frac{\delta\, {\cal A}_{\alpha}}{\delta\,\partial_{\mu}\chi_{\kappa}}\, {\widetilde{\cal A}}_{\alpha}^*\right)
+
{\cal A}_{\alpha}\,\frac{\delta\, {\widetilde{\cal A}}_{\alpha}^*}{\delta\,\chi_{\kappa}} -\partial_{\mu}\left({\cal A}_{\alpha}\,\frac{\delta\, {\widetilde{\cal A}}_{\alpha}^*}{\delta\,\partial_{\mu}\chi_{\kappa}}\right)
\label{topidentity}\\
&&
+\frac{\delta\, {\cal A}_{\alpha}^*}{\delta\,\chi_{\kappa}}\, {\widetilde{\cal A}}_{\alpha}-\partial_{\mu}\left(\frac{\delta\, {\cal A}_{\alpha}^*}{\delta\,\partial_{\mu}\chi_{\kappa}}\, {\widetilde{\cal A}}_{\alpha}\right)
+
{\cal A}_{\alpha}^*\,\frac{\delta\, {\widetilde{\cal A}}_{\alpha}}{\delta\,\chi_{\kappa}} -\partial_{\mu}\left({\cal A}_{\alpha}^*\,\frac{\delta\, {\widetilde{\cal A}}_{\alpha}}{\delta\,\partial_{\mu}\chi_{\kappa}}\right)
=0
\nonumber
\end{eqnarray}
By imposing the first order differential equations, or self-duality equations, on the fields as
\begin{equation}
{\cal A}_{\alpha}=\pm {\widetilde{\cal A}}_{\alpha}
\label{sdeqs}
\end{equation}
it follows that, together with the identities (\ref{topidentity}), they imply the equations 
\begin{eqnarray}
&&\frac{\delta\, {\cal A}_{\alpha}}{\delta\,\chi_{\kappa}}\, {{\cal A}}_{\alpha}^*-\partial_{\mu}\left(\frac{\delta\, {\cal A}_{\alpha}}{\delta\,\partial_{\mu}\chi_{\kappa}}\, {{\cal A}}_{\alpha}^*\right)
+
 {{\cal A}}_{\alpha}\,\frac{\delta\, {\cal A}_{\alpha}^*}{\delta\,\chi_{\kappa}}-\partial_{\mu}\left( {{\cal A}}_{\alpha}\,\frac{\delta\, {\cal A}_{\alpha}^*}{\delta\,\partial_{\mu}\chi_{\kappa}}\right)
 \label{eleqs}\\
&& +
\frac{\delta\, {\widetilde{\cal A}}_{\alpha}^*}{\delta\,\chi_{\kappa}}\,{\widetilde{\cal A}}_{\alpha} -\partial_{\mu}\left(\frac{\delta\, {\widetilde{\cal A}}_{\alpha}^*}{\delta\,\partial_{\mu}\chi_{\kappa}}\,{\widetilde{\cal A}}_{\alpha}\right)
+
{\widetilde{\cal A}}_{\alpha}^*\,\frac{\delta\, {\widetilde{\cal A}}_{\alpha}}{\delta\,\chi_{\kappa}} -\partial_{\mu}\left({\widetilde{\cal A}}_{\alpha}^*\,\frac{\delta\, {\widetilde{\cal A}}_{\alpha}}{\delta\,\partial_{\mu}\chi_{\kappa}}\right)=0
\nonumber
\end{eqnarray}
Note that (\ref{eleqs}) are the Euler-Lagrange equations associated to the functional 
\begin{equation}
E=\frac{1}{2}\,\int d^dx\,\left[{\cal A}_{\alpha}\,{\cal A}_{\alpha}^*+{\widetilde{\cal A}}_{\alpha}\,{\widetilde{\cal A}}_{\alpha}^*\right]
\label{energy}
\end{equation}
So, first order differential equations together with second order topological identities lead to second order Euler-Lagrange equations. Note that, if $E$ is positive definite then the self-dual solutions saturate a lower bound on $E$ as follows. From (\ref{sdeqs}) we have that ${\cal A}_{\alpha}^2={\widetilde{\cal A}}_{\alpha}^2=\pm {\cal A}_{\alpha}\,{\widetilde{\cal A}}_{\alpha}$. Note that (\ref{sdeqs}) also implies that ${\cal A}_{\alpha}\,{\widetilde{\cal A}}_{\alpha}^*={\cal A}_{\alpha}^*\,{\widetilde{\cal A}}_{\alpha}$. Therefore, if ${\cal A}_{\alpha}\,{\cal A}_{\alpha}^*\geq 0$, and consequently ${\widetilde{\cal A}}_{\alpha}\,{\widetilde{\cal A}}_{\alpha}^*\geq 0$, we have that 
\begin{eqnarray}
{\cal A}_{\alpha}= {\widetilde{\cal A}}_{\alpha}\quad &\rightarrow& \quad Q=\int d^dx\, {\cal A}_{\alpha}\,{\cal A}_{\alpha}^*\geq 0
\nonumber\\
{\cal A}_{\alpha}= -{\widetilde{\cal A}}_{\alpha}\quad &\rightarrow& \quad Q=-\int d^dx\, {\cal A}_{\alpha}\,{\cal A}_{\alpha}^*\leq 0
\end{eqnarray}
Therefore we have that 
\begin{equation}
E=\frac{1}{2}\,\int d^dx\,\left[{\cal A}_{\alpha} \mp {\widetilde{\cal A}}_{\alpha}\right] \left[{\cal A}_{\alpha}^* \mp {\widetilde{\cal A}}_{\alpha}^*\right]\pm \frac{1}{2}\,\int d^dx\,\left[{\cal A}_{\alpha} \, {\widetilde{\cal A}}_{\alpha}^*+{\cal A}_{\alpha}^* \, {\widetilde{\cal A}}_{\alpha}\right]\geq \mid Q\mid
\label{bound}
\end{equation}
and the equality holds true for self-dual solutions, where we have
\begin{equation}
E=\int d^dx\,{\cal A}_{\alpha}\,{\cal A}_{\alpha}^*=\int d^dx\,{\widetilde{\cal A}}_{\alpha}\,{\widetilde{\cal A}}_{\alpha}^*= \mid Q\mid
\end{equation}

The splitting  of the integrand of $Q$ as in (\ref{topcharge}) is quite arbitrary, but once it is chosen one can still change 
${\cal A}_{\alpha}$ and ${\widetilde{\cal A}}_{\alpha}$ by the apparently innocuous transformation 
\begin{equation}
{\cal A}_{\alpha}\rightarrow {\cal A}_{\alpha}^{\prime}={\cal A}_{\beta}\,k_{\beta\,\alpha}\;;\qquad
{\widetilde{\cal A}}_{\alpha}^* \rightarrow \left({\widetilde{\cal A}}_{\alpha}^{\prime}\right)^*=k^{-1}_{\alpha\,\beta}{\widetilde{\cal A}}_{\beta}^*
\label{newsplitting}
\end{equation}
The topological charge does not change and so it is still invariant under homotopic transformations. Therefore, we can now apply the same reasoning as above with the transformed quantities ${\cal A}_{\alpha}^{\prime}$ and ${\widetilde{\cal A}}_{\alpha}^{\prime}$. The transformed self-duality equations are
\begin{equation}
{\cal A}_{\beta}\,k_{\beta\,\alpha}=\pm \left(k^{-1}_{\alpha\,\beta}\right)^*{\widetilde{\cal A}}_{\beta} \quad\rightarrow \quad
{\cal A}_{\beta}\,h_{\beta\,\alpha}=\pm {\widetilde{\cal A}}_{\alpha}
\label{newsdeqs}
\end{equation}
where we have defined the hermitian and invertible matrix 
\begin{equation}
h\equiv k\,k^{\dagger}
\end{equation}
Together with the transformed identities (\ref{topidentity}), the new self-duality equations (\ref{newsdeqs}) imply the Euler-Lagrange equations associated to the energy
\begin{equation}
E^{\prime}=\frac{1}{2}\,\int d^dx\,\left[{\cal A}_{\alpha}\,h_{\alpha\beta}\,{\cal A}_{\beta}^*+{\widetilde{\cal A}}_{\alpha}\,h^{-1}_{\alpha\beta}\,{\widetilde{\cal A}}_{\beta}^*\right]
\label{newenergydual}
\end{equation}
Note that the matrix $h$, or equivalently $k$, can be used to introduce new fields in the theory without changing the topological charge $Q$ and therefore its field content. 

It is import to note that the new self-duality equations (\ref{newsdeqs}) will also imply the Euler-Lagrange equations, coming from $E^{\prime}$, associated to such new fields $h_{\alpha\beta}$. Indeed, if the topological charge does not depend upon these new fields,  so does not ${\cal A}_{\alpha}$ and ${\widetilde{\cal A}}_{\alpha}$. Then the Euler-Lagrange equations associated to the fields $h_{\alpha\beta}$ is 
\begin{equation}
{\cal A}_{\alpha}\,{\cal A}_{\beta}^*-{\widetilde{\cal A}}_{\gamma}\,h^{-1}_{\gamma\alpha}\,h^{-1}_{\beta\delta}\,{\widetilde{\cal A}}_{\delta}^*=0
\end{equation}
Note that such equations are implied  by the self-duality equations (\ref{newsdeqs}).

In addition, it follows that (\ref{newsdeqs}) implies ${\cal A}_{\alpha}\,h_{\alpha\beta}\,{\cal A}_{\beta}^*={\widetilde{\cal A}}_{\alpha}\,h^{-1}_{\alpha\beta}\,{\widetilde{\cal A}}_{\beta}^*=\pm {\cal A}_{\alpha}\,{\widetilde{\cal A}}_{\alpha}^*=\pm {\cal A}_{\alpha}^*\,{\widetilde{\cal A}}_{\alpha}$. Therefore, if ${\cal A}_{\alpha}\,h_{\alpha\beta}\,{\cal A}_{\beta}^*\geq 0$, and consequently ${\widetilde{\cal A}}_{\alpha}\,h^{-1}_{\alpha\beta}\,{\widetilde{\cal A}}_{\beta}^*\geq 0$, we have that the bound follows in the same way as before 
\begin{eqnarray}
E^{\prime}&=&\frac{1}{2}\,\int d^dx\,\left[{\cal A}_{\beta}\,k_{\beta\,\alpha}\mp \left(k^{-1}_{\alpha\,\beta}\right)^*{\widetilde{\cal A}}_{\beta} \right]\left[{\cal A}_{\gamma}^*\,k_{\gamma\,\alpha}^*\mp k^{-1}_{\alpha\,\gamma}{\widetilde{\cal A}}_{\gamma}^* \right]
\nonumber\\
&\pm& \frac{1}{2}\,\int d^dx\left[{\cal A}_{\alpha}\,{\widetilde{\cal A}}_{\alpha}^*+{\cal A}_{\alpha}^*\,{\widetilde{\cal A}}_{\alpha}\right]
\geq \mid Q\mid
\end{eqnarray}
We now discuss some examples where such  ideas have been applied. 

\section{Multi-Field Kinks in $(1+1)$-dimensions}
\label{sec:kinks}
\setcounter{equation}{0}

Self-dual sectors  for theories in $(1+1)$-dimensions, containing just one scalar field, like the sine-Gordon, and $\lambda\,\phi^4$ models, have been known for quite a long time. The application of the ideas explained in Section \ref{sec:introduction} have lead to the construction of self-dual sectors in theories containing any number of scalar fields in $(1+1)$-dimensions \cite{sd2d,bpscomments}. We consider here theories of real scalar fields. In such case, the relevant topological charge is given by 
\begin{equation}
Q=\int_{-\infty}^{\infty}dx\, \frac{d\, U}{d\,x}= \int_{-\infty}^{\infty}dx\,\frac{\delta\, U}{\delta\,\varphi_a}\,\frac{d\,\varphi_a}{d\,x}=
U\left(\varphi_a(x=\infty)\right)-U\left(\varphi_a(x=-\infty)\right).
\label{topcharge2d}
\end{equation}
where $U$ is an arbitrary real functional of the real scalar fields $\varphi_a$, $a=1,2,\ldots r$, but not of their derivatives. Clearly, the density of such a topological charge has the form given in (\ref{topcharge}), and following (\ref{newsplitting}) we can split it as (the quantities ${\cal A}_{\alpha}$ and ${\tilde {\cal A}}_{\alpha}$ are real, and so is the matrix $k$)
\begin{equation}
{\cal A}_{\alpha}\equiv k_{ab}\, \frac{d\,\varphi_b}{d\,x}\;;\qquad\qquad\qquad {\tilde {\cal A}}_{\alpha}\equiv \frac{\delta\, U}{\delta\,\varphi_b}\,k^{-1}_{ba}\,, 
\label{choiceofacal}
\end{equation}
where $k_{ab}$ is an arbitrary invertible matrix that can be introduced due to the freedom in the splitting. According to (\ref{newsdeqs}),  the self-duality equations are
\begin{equation}
 \eta_{ab}\,\frac{d\,\varphi_b}{d\,x}=\pm \,  \frac{\delta\, U}{\delta\,\varphi_a},\qquad\qquad\qquad\qquad \eta= k^T\,k
 \label{bpseq}
 \end{equation}
 and so, $\eta_{ab}$ is an invertible symmetric matrix. In what follows, we shall take $\eta_{ab}$ to be a constant matrix, and not a matrix containing new fields, as it is allowed by the construction discussed in Section \ref{sec:introduction}. It will play the role of a metric in the target space of the scalar fields $\varphi_a$.  
 
 Following (\ref{newenergydual}) the static energy of our theory is 
 \begin{equation}
 E= \int_{-\infty}^{\infty} dx\,\left[ \frac{1}{2}\,\eta_{ab}\, \frac{d\,\varphi_a}{d\,x}\, \frac{d\,\varphi_b}{d\,x} + V\right],
 \label{energyfunc}
 \end{equation}
 where  the potential is given by
 \begin{equation}
 V=\frac{1}{2}\,\eta^{-1}_{ab}\,\frac{\delta\, U}{\delta\,\varphi_a}\,\frac{\delta\, U}{\delta\,\varphi_b}
 \label{potdef}
 \end{equation}
 Therefore, from the arguments of Section \ref{sec:introduction}, it follows that solutions of (\ref{bpseq}) are solutions of the static Euler-Lagrange equations associated to the energy functional (\ref{energyfunc}). The quantity $U$ plays the role of a pre-potential. Note that given a choice of pre-potential $U$ one can directly obtain the potential $V$ and so a scalar field theory with a self-dual sector. However, given the potential $V$ it is not in general easy to find the pre-potential $U$. We shall discuss here the construction of self-dual theories from the choice of pre-potential. 

We restrict our discussion to the cases where the scalar fields $\varphi_a$, the pre-potential $U$, and the matrix $\eta_{ab}$ are real. In addition, we are interested in the cases for which the static energy functional $E$, given in (\ref{energyfunc}), is positive definite. Thus we need to restrict our discussion to cases in which all the eigenvalues of $\eta_{ab}$ are positive definite. In order for  the self-dual solutions of (\ref{bpseq}) to possess finite energy $E$,  we need the energy density to vanish at spatial infinities when evaluated on such solutions, and so, given our restrictions, we require that   
\begin{equation}
\frac{d\,\varphi_b}{d\,x}\rightarrow 0\;;\qquad\qquad   \frac{\delta\, U}{\delta\,\varphi_a}\rightarrow 0\;;\qquad\qquad {\rm as}\qquad x\rightarrow \pm \infty.
\label{vacuaconditions}
\end{equation}
Thus, the self-duality equations (\ref{bpseq}) should possess constant vacua solutions $\varphi_a^{\rm (vac.)}$ that are zeros of all the first derivatives of the pre-potential, {\it i.e.} 
\begin{equation}
\frac{\delta\, U}{\delta\,\varphi_b}\mid_{\varphi_a=\varphi_a^{\rm (vac.)}}=0.
\label{vacuau}
\end{equation} 
We then see from (\ref{potdef}) that such vacua are also zeros of the potential $V$ and of its first derivatives, {\it i.e.}
\begin{equation}
V\left(\varphi_a^{\rm (vac.)}\right)=0\;;\qquad\qquad\qquad 
\frac{\delta\, V}{\delta\,\varphi_b}\mid_{\varphi_a=\varphi_a^{\rm (vac.)}}=0.
\end{equation}
Moreover, we would like the theories we are constructing to possess various soliton type solutions, and we know that, in general, the total topological charges of such solutions are obtained by additions, under some finite or infinite abelian group, of the charges of the constituent one-solitons. Thus, we would like to have systems of vacua as degenerate as possible. Certainly there are numerous ways  of achieving this goal. We shall use a group theoretical approach to the construction of  the pre-potentials $U$, as we now explain.

 Consider a Lie algebra ${\cal G}$ and let ${\vec \alpha}_a$, $a=1,2,\ldots r\equiv {\rm rank}\,{\cal G}$, be the set of its simple roots.  We use the scalar fields $\varphi_a$ to construct our basic vector in the root space: 
\begin{equation}
 {\vec \varphi}\equiv \sum_{a=1}^r \varphi_a\, \frac{2\,{\vec \alpha}_a}{{\vec \alpha}_a^2}.
 \label{phivectorb}
 \end{equation}
Next we choose a representation  ${\cal R}$ (irreducible or not) of the Lie algebra ${\cal G}$, and we denote by ${\vec \mu}_k$  the set of weights of ${\cal R}$. We take the pre-potential $U$ to be of the form
\begin{equation}
U\equiv \sum_{{\vec \mu}_k\in{\cal R}^{(+)}}\left[ \gamma_{{\vec \mu}_k}\, \cos\left({\vec \mu}_k\cdot {\vec \varphi}\right)+
\delta_{{\vec \mu}_k}\, \sin\left({\vec \mu}_k\cdot {\vec \varphi}\right) \right],
\label{prepotconstruct2}
\end{equation}
where the superscript $+$ in ${\cal R}^{(+)}$  means that if ${\cal R}$ possesses pair of weights of the form $\left({\vec \mu}_k\,,\,-{\vec \mu}_k\right)$, we take just one member of the pair. There are several ways of having the pre-potential (\ref{prepotconstruct2}) satisfying (\ref{vacuau}), and so the vacuum structure of our theories can be quite complicated. For details see \cite{sd2d}. In order to clarify the aspects of the construction we discuss here and example for the $SU(3)$ group. 

\subsection{An $SU(3)$ example}

The rank of $SU(3)$ is two and so we have two fields, $\varphi_1$ and $\varphi_2$, in this case. We take the matrix $\eta_{ab}$ to be of the form
\begin{eqnarray}
\eta=\left(
\begin{array}{cc}
2&-\,\lambda\\
-\,\lambda &2
\end{array}\right),
\qquad\qquad\qquad
\eta^{-1}=\frac{1}{4-\lambda^2}\left(
\begin{array}{cc}
2&\lambda\\
\lambda&2
\end{array}\right),
\label{etamatricessu3lambda}
\end{eqnarray}
where we have introduced a real parameter $\lambda$. The eigenvalues of $\eta$ are $2\pm \lambda$, and so we have to keep $\lambda$ in the interval $-2<\lambda<2$, to have $\eta$ positive definite and invertible.  The weights of the triplet representation of $SU(3)$ are given by
\begin{equation}
{\vec \mu}_1={\vec \lambda}_1, \qquad\qquad {\vec \mu}_2={\vec \lambda}_1-{\vec \alpha}_1,
\qquad\qquad {\vec \mu}_3={\vec \lambda}_1-{\vec \alpha}_1-{\vec \alpha}_2
\end{equation}
where  $\alpha_a$, $a=1,2$ are the simple roots of $SU(3)$, and ${\vec \lambda}_1$, is the fundamental weights which is the highest weight of the triplet representation.  Thus  from (\ref{prepotconstruct2}) we get the pre-potential as
\begin{equation}
U=\gamma_1\,\cos \varphi_1+\gamma_2\,\cos\varphi_2+\gamma_3\,\cos\left(\varphi_1-\varphi_2\right),
\label{usu3tripletantitriplet}
\end{equation}
where we have chosen the $\delta$-terms in (\ref{prepotconstruct2})  to vanish. The static energy (\ref{energyfunc}) now becomes  
\begin{equation}
E=\int_{-\infty}^{\infty}dx\,\left[\left(\partial_x\varphi_1\right)^2+\left(\partial_x\varphi_2\right)^2-\lambda\,\partial_x\varphi_1\,\partial_x\varphi_2 + V\left(\varphi_1,\varphi_2\right) \right],
\label{energysu3}
\end{equation}
where the potential (\ref{potdef}) is given by
\begin{eqnarray}
V=
&&\left[-\gamma_1^2 \sin ^2(\varphi_1)+\gamma_1 \sin (\varphi_1) (\gamma_3
   (\lambda-2) \sin (\varphi_1-\varphi_2)
      \right. 
    \nonumber\\
   &&-\left. \gamma_2^2 \sin ^2(\varphi_2)-\gamma_2 \gamma_3 (\lambda-2) \sin
   (\varphi_2) \sin (\varphi_1-\varphi_2)
  \right. 
  \label{vsu3tripletantitriplet} \\
   &&-\left. 
   \gamma_2 \lambda \sin
   (\varphi_2))
+\gamma_3^2 (\lambda-2) \sin
   ^2(\varphi_1-\varphi_2)\right]/\left(\lambda^2-4\right)
  \nonumber
   \end{eqnarray}

 The self-duality equations (\ref{bpseq}) are  now of the form:
   \begin{eqnarray}
   \partial_x\varphi_1&=&\pm 
   \frac{\left[2 \gamma_1 \sin (\varphi_1)+\gamma_2 \lambda \sin (\varphi_2)-\gamma_3
   (\lambda-2) \sin (\varphi_1-\varphi_2)\right]}{\lambda^2-4},
   \label{su3bpseq33bar}\\
   \partial_x\varphi_2&=&\pm 
   \frac{\left[\gamma_1 \lambda \sin (\varphi_1)+2 \gamma_2 \sin (\varphi_2)+\gamma_3
   (\lambda-2) \sin (\varphi_1-\varphi_2)\right]}{\lambda^2-4}.
   \nonumber
   \end{eqnarray}
   The vacua are determined by the conditions (\ref{vacuau}) which in this case become
    \begin{eqnarray}
\frac{\partial U}{\partial \varphi_1}\mid_{\varphi_a=\varphi_a^{\rm (vac.)}}&=&   
   -\gamma_1 \sin (\varphi_1^{\rm (vac.)})-\gamma_3 \sin (\varphi_1^{\rm (vac.)}-\varphi_2^{\rm (vac.)})=0,
   \label{vacuasu3}\\
   \frac{\partial U}{\partial \varphi_2}\mid_{\varphi_a=\varphi_a^{\rm (vac.)}}&=&  
   \gamma_3 \sin (\varphi_1^{\rm (vac.)}-\varphi_2^{\rm (vac.)})-\gamma_2 \sin (\varphi_2^{\rm (vac.)})=0,
   \nonumber
   \end{eqnarray}
  and these conditions imply that
   \begin{equation}
   \gamma_1 \sin (\varphi_1^{\rm (vac.)})=-\gamma_3 \sin (\varphi_1^{\rm (vac.)}-\varphi_2^{\rm (vac.)})=-\gamma_2 \sin (\varphi_2^{\rm (vac.)}).
   \label{vacuasu3b}
   \end{equation}
Certainly (\ref{vacuasu3b}) are satisfied if 
\begin{equation}
\varphi_a^{\rm (vac.)}=\pi\,n_a;\quad n_a\in \relax\leavevmode\hbox{\sf Z\kern-.4em Z}; \quad  a=1,2;\quad 
\mbox{\rm any values of the $\gamma$'s}\label{firstkindvacuasu3}
\end{equation}
 However, we also have the additional vacua, depending   upon the particular values of the $\gamma$-constants that we are free to choose. For instance, one finds that  (\ref{vacuasu3b}) are satisfied if
\begin{eqnarray}
\left(\varphi_1^{\rm (vac.)}\,,\, \varphi_2^{\rm (vac.)}\right)&=&\left(\frac{2\pi}{3}+2\,\pi\,n_1\,,\, \frac{4\pi}{3}+2\,\pi\,n_2\right)\;;\qquad \qquad \gamma_1=\gamma_2=\gamma_3=1,
\nonumber\\ 
\left(\varphi_1^{\rm (vac.)}\,,\, \varphi_2^{\rm (vac.)}\right)&=&\left(\frac{4\pi}{3}+2\,\pi\,n_1\,,\, \frac{2\pi}{3}+2\,\pi\,n_2\right)\;;\qquad \qquad n_1\,,\,n_2\in \relax\leavevmode\hbox{\sf Z\kern-.4em Z}.\label{othervacuasu3}
\end{eqnarray}

\subsection{A mechanical interpretation of the self-dual solutions}

As we have seen in (\ref{vacuaconditions}) and (\ref{vacuau}), the finite energy solutions of the self-duality equations (\ref{bpseq}) have to go to  constant vacua solutions for $x\rightarrow \pm \infty$.  Therefore, each of these solutions connect two vacua of the theory. In order to have a geometric picture of these solutions let us write the self-duality equations (\ref{bpseq}) as 
\begin{equation}
{\vec v}=\pm {\vec \nabla}_{\eta}U\;; \qquad\qquad {\rm with}\qquad \left({\vec v}\right)_a =\frac{d\,\varphi_a}{d\,x}\;;\qquad 
\left({\vec \nabla}_{\eta}U\right)_a= \eta^{-1}_{ab}\frac{\delta\, U}{\delta\,\varphi_b}.
 \label{equiv_bpseq}
 \end{equation}

 Given the pre-potential $U$ and the metric $\eta_{ab}$, which we assume real, constant and positive definite, the $\eta$-gradient of $U$ defines curves in the space of $\varphi_1,\ldots,\varphi_r$, with ${\vec \nabla}_{\eta}U$ being the tangent vector to these curves. The curves never intersect each other, since otherwise ${\vec \nabla}_{\eta}U$ would not be uniquely defined on a given point in $\varphi$-space.  They can at most touch each other tangentially, or meet at points where ${\vec \nabla}_{\eta}U$ vanishes. The self-duality equation is a first order partial differential equation and so a given solution is determined by the values of the fields $\varphi_a$ at a given point $x=x_0$.
 
 The geometric picture is therefore that of a particle traveling in the $\varphi$-space with $x$-velocity ${\vec v}$, and with the space coordinate $x$ playing the role of time. Therefore, the problem of solving the self-duality equation (\ref{bpseq}) reduces to that of constructing the curves in the $\varphi$-space determined by the $\eta$-gradient of $U$. Any particular solution corresponds to a particular curve  determined by the initial values $\varphi_a\left(x_0\right)$. The finite energy solutions correspond to the curves that start and end at the extrema of the pre-potential $U$, {\it i.e.} at the points where ${\vec \nabla}_{\eta}U$ vanishes. 

Consider now a given curve $\gamma$ in the $\varphi$-space, parameterized by $x$, {\it i.e.} $\varphi_a\left(x\right)$, which is a solution of the self-duality equation (\ref{bpseq}), and associated to this curve define the quantity
\begin{equation}
{\cal Q}\left(\gamma\right)=\int_{\gamma}dx\,{\vec v}\cdot {\vec \nabla}U= \int_{\gamma}dx\,\frac{d\,\varphi_a}{d\,x}\,\frac{\delta\,U}{\delta\,\varphi_a}=U\left(x_f\right)-U\left(x_i\right),
\end{equation}
where $x_f$ and $x_i$ correspond to the final and initial points respectively, of the curve $\gamma$. Note that the tangent vector to this curve is ${\vec \nabla}_{\eta}U$ and not the ordinary gradient of $U$, {\it i.e.} ${\vec \nabla}U$, since the curve is a solution of the self-duality equations (\ref{bpseq}). From these self-duality equations we see that
\begin{equation}
{\cal Q}\left(\gamma\right)=\pm \int_{\gamma}dx\,\eta_{ab}\,\frac{d\,\varphi_a}{d\,x}\,\frac{d\,\varphi_b}{d\,x}=\pm \int_{\gamma}dx\,\omega_a\left(\frac{d\,{\tilde \varphi}_a}{d\,x}\right)^2,
\label{qgammadef}
\end{equation}
where we have diagonalized the matrix $\eta$, {\it i.e.}
\begin{equation}
\eta=\Lambda^{T}  \,\eta^D\, \Lambda\;;\qquad\qquad\Lambda^T\,\Lambda=\hbox{{1}\kern-.25em\hbox{l}}\;; \qquad\qquad \eta^D_{ab}=\omega_a \,\delta_{ab}\;;\qquad\qquad \omega_a>0
\label{diagonaletadef}
\end{equation}
and have assumed that the eigenvalues of $\eta$ are all positive, and have defined ${\tilde \varphi}_a=\Lambda_{ab}\,\varphi_b$. Under the assumption that $\eta$ is positive definite, one observes that ${\cal Q}\left(\gamma\right)$ can only vanish  if the fields are constant along the whole curve, or in other words, if the curve is just a point. Therefore, the solutions of the self-duality equations cannot start and end on points in the $\varphi$-space, where the the pre-potential $U$ has the same value.  In fact, there is more to this. As one progresses along the curve, the difference between the value of the  pre-potential $U$ at this particular point and at the initial point, only increases in modulus. This means that the curve, that is a solution of the self-duality equations (\ref{bpseq}), climbs the pre-potential $U$, either upwards or downwards, without ever returning to an altitude that it has already passed through.  

\subsection{A connection with Hamilton-Jacobi equation}

For a mechanical system with Hamiltonian $H=H\left(\varphi_a\,,\,p_a\,,\,t\right)$, where $\varphi_a$ and $p_a$ are the canonical coordinates in phase space, the Hamilton-Jacobi equation is given by
\begin{equation}
H\left(\varphi_a\,,\,\frac{\partial\,S}{\partial\,\varphi_a}\,,\,t\right) +\frac{\partial\,S}{\partial\,t}=0
\label{properhj}
\end{equation}
with $S$ being Hamilton's principal function, which is related to the momenta by
\begin{equation}
p_a=\frac{\partial\, S}{\partial\,\varphi_a}
\end{equation}
The Euler-Lagrange equations associated to the static energy functional (\ref{energyfunc}) are given by
\begin{equation}
\eta_{ab}\,\partial_x^2\varphi_b=\frac{\partial \, V}{\partial\,\varphi_a}\qquad\qquad\qquad\qquad a,b=1,2,\ldots r
\label{staticeq}
\end{equation}
One can interpret such an equation as the Newton equation for a particle, of unit mass, moving in a $r$-dimensional $\varphi$-space with metric $\eta$, with time being the $x$ coordinate, and under the action of an inverted potential, i.e. 
\begin{equation}
 {\tilde V}\rightarrow -V\qquad\qquad\qquad x\rightarrow t
 \end{equation}
Identifying the  pre-potential $U$ with Hamilton's principal function up to a sign, i.e. $S\equiv \pm U$, one then gets the relation (\ref{potdef}) can be written as
\begin{equation}
 \frac{1}{2}\,\eta^{-1}_{ab}\,\frac{\delta\, S}{\delta\,\varphi_a}\,\frac{\delta\, S}{\delta\,\varphi_b}+{\tilde V}=0
 \label{potdefhamiltonjacobi}
 \end{equation}
 But that is just the Hamilton-Jacobi equation (\ref{properhj}) for the Hamiltonian
 \begin{equation}
H=\frac{1}{2}\,\eta_{ab}^{-1}\,p_a\,p_b + {\tilde V}\left(q\right)
\label{hamhj}
\end{equation}
as we are assuming the that $U$, and so $S$, does not depend upon time, i.e. $x$. 
In its turn, the self-duality equations (\ref{bpseq}) become just the kinematical relation between momenta and velocities 
\begin{equation}
{\dot \varphi}_a=\eta^{-1}_{ab}\, p_b
\label{bpshj}
\end{equation}
The Lagragian associated to the Hamiltonian (\ref{hamhj}) is
\begin{equation}
L=\frac{1}{2}\,\eta_{ab}\,{\dot \varphi}_a\,{\dot \varphi}_b - {\tilde V}\left(\varphi\right)
\label{laghj}
\end{equation}
Note also that the topological charge (\ref{topcharge2d})  takes the form of an action 
\begin{equation}
Q=\pm \int_{-\infty}^{\infty}dt\, p_a\, {\dot \varphi}_a 
\label{topchargehj}
\end{equation}
The Hamilton-Jacobi equation (\ref{potdefhamiltonjacobi}) (see (\ref{properhj})) implies that the Hamiltonian vanishes, i.e.
\begin{equation}
H=0 \qquad\qquad \rightarrow \qquad\qquad 
-2\, {\tilde V} = \eta_{ab}^{-1}\,p_a\,p_b= p_a\, {\dot \varphi}_a
\end{equation}
and so we can write
\begin{equation}
Q=\pm \int_{-\infty}^{\infty}dt\, \left[p_a\, {\dot \varphi}_a -H\right]
\label{topchargehj2}
\end{equation}
The static energy (\ref{energyfunc}) becomes
\begin{eqnarray}
 E=\int_{-\infty}^{\infty}dt\, \left[\frac{1}{2}\,\eta_{ab}\,{\dot \varphi}_a\,{\dot \varphi}_b - {\tilde V}\left(\varphi\right) \right]=\int_{-\infty}^{\infty}dt\, \left[ H - 2\, {\tilde V}\left(\varphi\right) \right]
 = \pm Q
\end{eqnarray}
Therefore, we have a mechanical system of a particle in $d$ dimensions, and the BPS solutions correspond to solutions of  such a system where the energy, measured by $H$, vanishes. The Hamilton-Jacobi equation leads to the equality of the static energy, measured by $E$, to  the topological charge $Q$. The BPS equation itself is just the  kinematical relation between velocities and momenta. 

\section{Lumps in $(2+1)$-dimensions}
\label{sec:lumps}
\setcounter{equation}{0}

As an example of a theory with self-dual sector we shall consider the $CP^{N-1}$ model in $(2+1)$-dimensions. $CP^{N-1}$ is the $N-1$ dimensional complex projective space, i.e. the space of all equivalent classes of complex vectors $z=\left(z_1\,,\,z_2\,,\,\ldots z_{N}\right)$, such that two vectors $z$ and $z^{\prime}$ are equivalent if $z^{\prime}=\lambda\,z$, with $\lambda$ being a complex number \cite{dadda,wojtekbook}. We shall take the representatives of such classes to be the unit vectors
\begin{equation}
z=\left(z_1\,,\,z_2\,,\,\ldots z_{N}\right)\qquad\qquad\qquad z_a^*\,z_a=1
\end{equation}
$CP^{N-1}$ is isomorphic to the hermitian symmetric space $SU(N)/SU(N-1)\otimes U(1)$. Indeed, $SU(N)$ acts transitively on the vectors $z$ through its defining $N$-dimensional representation. As such representation is unitary its action preserves the modulus of the vectors $z$, and a given vector, let us say $z=\left(0\,,\, 0\,,\, \ldots 1\right)$, is left invariant by $(N-1)\times (N-1)$ unitary matrices, i.e. the subgroup $U(N-1)=SU(N-1)\otimes U(1)$.  The isometry subgroup of any other vector $z$ is isomorphic to $SU(N-1)\otimes U(1)$. 

The second homotopy group of $CP^{N-1}$ is isomorphic to the integers under addition, i.e. $\pi_2\left(SU(N)/SU(N-1)\otimes U(1)\right)=\relax\leavevmode\hbox{\sf Z\kern-.4em Z}$. The corresponding topological charge has an integral representation given by 
\begin{equation}
Q=\frac{1}{2\,\pi}\,\int d^2x\, \varepsilon_{\mu\nu}\, \partial_{\mu}\,A_{\nu}
\label{topchargecpn}
\end{equation}
with
\begin{equation}
A_{\mu}=\frac{i}{2}\left(z^{\dagger}\,\partial_{\mu}z-\partial_{\mu}z^{\dagger}\,z\right)
\end{equation}
and where the integration in (\ref{topchargecpn}) is on the two dimensional plane $\left(x_1\,,\,x_2\right)$, which by identifying the spatial infinity becomes isomorphic to $S^2$. 
Under the local phase transformation $z\rightarrow e^{i\alpha}\,z$, we have that $A_{\mu}\rightarrow A_{\mu}-\partial_{\mu}\alpha$. Introducing the covariant derivative $D_{\mu}\equiv \partial_{\mu}+i\,A_{\mu}$, we can write (\ref{topchargecpn}) as
\begin{equation}
Q=\frac{i}{2\,\pi}\,\int d^2x\, \varepsilon_{\mu\nu}\, \left(D_{\mu}z\right)^{\dagger}\,D_{\nu}z
=\frac{1}{4\,\pi}\,\int d^2x\left[  \left(D_{\mu}z\right)^{\dagger}\,i\,\varepsilon_{\mu\nu}\,D_{\nu}z+\left(i\,\varepsilon_{\mu\nu}\,D_{\nu}z\right)^{\dagger}\, D_{\mu}z
\right]
\label{topchargecpn2}
\end{equation}
Following (\ref{newsplitting}) we define the quantities 
\begin{equation}
{\cal A}_{\mu}^a=\left(D_{\mu}z\right)_b\,k_{ba}\qquad\qquad\qquad 
{\widetilde{\cal A}}_{\mu}^a=\left(k_{ab}^{-1}\right)^*\, i\,\varepsilon_{\mu\nu}\,\left(D_{\nu}z\right)_b
\end{equation}
and so, the charge (\ref{topchargecpn2}) can be written in the form (\ref{topcharge}). From (\ref{newsdeqs}) the self-duality equations are given by
\begin{equation}
\left(D_{\mu}z\right)_b\,h_{ba}=\pm i\,\varepsilon_{\mu\nu}\,\left(D_{\nu}z\right)_a
\label{sdeqscpn}
\end{equation}
According to (\ref{newenergydual}), the energy functional becomes
\begin{equation}
E=\frac{1}{2}\,\int d^2x\left[ \left(D_{\mu}z\right)_a^*\,h_{ab}\,\left(D_{\mu}z\right)_b+\left(D_{\mu}z\right)_a^*\,h_{ab}^{-1}\,\left(D_{\mu}z\right)_b\right]
\label{energycpngeneral}
\end{equation}
Note however, that by contracting both sides of (\ref{sdeqscpn}) with $\varepsilon_{\rho\mu}$, one gets
\begin{equation}
\left(D_{\mu}z\right)_a=\pm i\,\varepsilon_{\mu\nu}\,\left(D_{\nu}z\right)_b\,h_{ba}
\label{sdeqscpn2}
\end{equation}
Therefore, (\ref{sdeqscpn}) and (\ref{sdeqscpn2}) imply
\begin{equation}
\left(D_{\mu}z\right)_b\left(h_{ba}-h^{-1}_{ba}\right)=0\qquad\rightarrow \qquad h^2=\hbox{{1}\kern-.25em\hbox{l}}
\end{equation}
But an hermitian matrix can be diagonalized by an unitary transformation, $h=U\,h_D\,U^{\dagger}$, with $h_D$ diagonal. Therefore
\begin{equation}
h^2=\hbox{{1}\kern-.25em\hbox{l}} \qquad\rightarrow \qquad h_D^2=\hbox{{1}\kern-.25em\hbox{l}}
\end{equation}
and so the square of the eigenvalues of $h$ have to be unity, i.e. $\lambda_a^2=1$. But in order for the energy $E$, given in (\ref{energycpngeneral}), to be positive definite, we need all the eigenvalues of $h$ to have the same sign. Consequently, we have that 
\begin{equation}
h=\hbox{{1}\kern-.25em\hbox{l}}
\label{honecpn}
\end{equation}
In such case, (\ref{sdeqscpn}) reduces to the self-duality equation for the usual  $CP^{N-1}$ model \cite{dadda,wojtekbook} 
\begin{equation}
\left(D_{\mu}z\right)_a=\pm i\,\varepsilon_{\mu\nu}\,\left(D_{\nu}z\right)_a
\label{sdeqscpn3}
\end{equation}
and (\ref{energycpngeneral}) to the energy of the usual  $CP^{N-1}$ model
\begin{equation}
E=\int d^2x \left(D_{\mu}z\right)^{\dagger}\,D_{\mu}z
\label{energycpn}
\end{equation}
In order to construct the self-dual solutions for (\ref{sdeqscpn3}), it is better to introduce the complex fields $u_a$ as
\begin{equation}
\left(u_1\,,\,u_2\,,\,\ldots u_{N-1}\,,\, 1\right)=\frac{1}{z_N}\,\left(z_1\,,\, z_2\ldots \,,\, z_{N-1}\,,\, z_N\right)
\end{equation}
One could have divided the vector of complex fields $z$, by any other component $z_a$, and the construction would be equivalent. It then follows that the covariant derivative becomes
\begin{eqnarray}
\left(D_{\mu}z\right)_{\alpha}&=&z_N\, \Omega_{\alpha\beta}\,\partial_{\mu}u_{\beta} ; \qquad \alpha\,,\,\beta=1,2,\ldots N-1
\nonumber\\
\left(D_{\mu}z\right)_{N}&=&-z_N\,\frac{u^{\dagger}\partial_{\mu}u}{1+u^{\dagger}\,u}
\end{eqnarray}
with
\begin{equation}
\Omega_{\alpha\beta}=\delta_{\alpha\beta}-\frac{u_{\alpha}\,u^*_{\beta}}{1+u^{\dagger}\,u}
; \qquad \qquad\qquad\alpha\,,\,\beta=1,2,\ldots N-1
\end{equation}
Therefore, the self duality equations (\ref{sdeqscpn3}) become
\begin{eqnarray}
\left[D_{\mu}z\mp i\,\varepsilon_{\mu\nu}\,D_{\nu}z\right]_{\alpha}&=&z_N\, \Omega_{\alpha\beta}\left[\partial_{\mu}u_{\beta}\mp i\,\varepsilon_{\mu\nu}\,\partial_{\nu}u_{\beta}\right]; \qquad \qquad\qquad\alpha\,,\,\beta=1,2,\ldots N-1
\nonumber\\
\left[D_{\mu}z\mp i\,\varepsilon_{\mu\nu}\,D_{\nu}z\right]_{N}&=&-z_N\,\frac{u^*_{\beta}}{1+u^{\dagger}\,u}\left[\partial_{\mu}u_{\beta}\mp i\,\varepsilon_{\mu\nu}\,\partial_{\nu}u_{\beta}\right]
\end{eqnarray}
Therefore,  the self duality equations (\ref{sdeqscpn3}) imply that 
\begin{equation}
\partial_{\mu}u_{\alpha}=\pm i\,\varepsilon_{\mu\nu}\,\partial_{\nu}u_{\alpha};\qquad\qquad\qquad \alpha=1,2,\ldots N-1
\end{equation}
These are Cauchy-Riemann equations for the $u$-fields. Indeed, the upper sign $(+)$ implies that $u$ is holomorphic, i.e. $u_{\beta}=u_{\beta}\left(w\right)$, and the lower sign $(-)$  that $u$ is anti-holomorphic, i.e. $u_{\beta}=u_{\beta}\left(w^*\right)$, where $w=x_1+i\,x_2$.

\section{Monopoles in $(3+1)$-dimensions}
\label{sec:monopoles}
\setcounter{equation}{0}

We now consider the case of the topological magnetic charge  defined by the integral over the three dimensional space $\relax\leavevmode\hbox{\rm I\kern-.18em R}^3$ 
\begin{equation}
Q_{M}=-\frac{1}{2}\int_{\relax\leavevmode\hbox{\rm I\kern-.18em R}^3} d^3x\,\varepsilon_{ijk}{\rm Tr}\left(F_{ij}\,D_k\Phi\right)=\int_{\relax\leavevmode\hbox{\rm I\kern-.18em R}^3} d^3x\,{\rm Tr}\left(B_{i}\,D_i\Phi\right)
\label{topchargeymhintro}
\end{equation}
where $B_i=-\frac{1}{2}\,\varepsilon_{ijk}\,F_{jk}$ is the non-abelian magnetic field, $F_{ij}=\partial_iA_j-\partial_jA_i+i\,e\,\left\lbrack\,{A_i}\, ,\,{A_j}\,\right\rbrack=F_{ij}^a\,T_a$, is the field tensor, $A_i=A_i^a\,T_a$, the gauge field, and $\Phi=\Phi_a\,T_a$, the Higgs field in the adjoint representation of a simple, compact, Lie group $G$, with generators $T_a$, $a=1,2,\ldots {\rm dim}\,G$. In addition, $D_i *=\partial_i*+i\,e\,\left\lbrack\,{A_i}\, ,\,{*}\,\right\rbrack$ is the covariant derivative in the adjoint representation of $G$. 

In this case all the fields are real and so, following (\ref{newsdeqs}) and the results of \cite{henrique}, we introduce the real quantities
\begin{equation}
{\cal A}_{\alpha}\equiv B_i^b\,k_{ba}\;;\qquad {\widetilde{{\cal A}}_{\alpha}}\equiv k^{-1}_{ab}\,\left(D_i\Phi\right)^b
\end{equation}
and so (\ref{topchargeymhintro}) can be written as in (\ref{topcharge}). The self-duality equations (\ref{newsdeqs}) become 
\begin{equation}
\frac{1}{2}\,\varepsilon_{ijk}\,F_{jk}^b\,h_{ba}=\pm \left(D_i\Phi\right)^a\qquad\qquad\qquad h=k\,k^T
\label{sdeqsymhintro}
\end{equation}
with $h_{ab}$, $a\,,\, b=1,2,\ldots {\rm dim}\,G$,  a symmetric invertible matrix of scalar fields. The equations (\ref{sdeqsymhintro})  constitute a generalization of the so-called BPS (Bogomolny-Prasad-Sommerfiled) equations \cite{bogo,prasad} for self-dual monopoles. The energy functional (\ref{newenergydual}) becomes \cite{henrique} 
\begin{equation}
E_{YMH}=\int d^3x\,\left[\frac{1}{4}\,h_{ab}\,F_{ij}^a\,F_{ij}^b+\frac{1}{2}\,h^{-1}_{ab}\,\left(D_i\Phi\right)^a\,\left(D_i\Phi\right)^b\right]
\label{energyymhintro}
\end{equation}
We then have a theory with gauge fields $A_{\mu}$, Higgs field $\Phi$ in the adjoint representation of the gauge group $G$, and $\left[{\rm dim}\,G\left({\rm dim}\,G+1\right)/2\right]$ real scalars fields assembled in the real, symmetric and invertible matrix $h$. The self-duality equations (\ref{sdeqsymhintro})  imply not only the static Euler-Lagrange equations associated to the gauge and Higgs fields, but also the ones associated to the scalar fields $h_{ab}$. 

The energy (\ref{energyymhintro}) evaluated on the self-dual solutions of (\ref{sdeqsymhintro}) is equal to the magnetic charge 
\begin{equation}
E_{YMH}= Q_M
\label{energytopchargerel}
\end{equation}

Under a gauge transformation $A_{\mu}\rightarrow g\,A_{\mu}\,g^{-1}+\frac{i}{e}\,\partial_{\mu}g\,g^{-1}$, we have that
$F_{\mu\nu}\rightarrow g\,F_{\mu\nu}\,g^{-1}$ and $D_{\mu}\Phi\rightarrow g\,D_{\mu}\Phi\,g^{-1}$. Therefore, energy (\ref{energyymhintro}) and the self-duality equations (\ref{sdeqsymhintro}) are invariant under
\begin{eqnarray}
F_{\mu\nu}^a&\rightarrow& d_{ab}\left(g\right)\,F_{\mu\nu}^b\,;\qquad 
\left(D_{\mu}\Phi\right)^a\rightarrow d_{ab}\left(g\right)\,\left(D_{\mu}\Phi\right)^b
\nonumber\\
h_{ab}&\rightarrow& d_{ac}\left(g\right)\,d_{bd}\left(g\right)\,h_{cd}
\label{gaugetransformgen}
\end{eqnarray}
where $d\left(g\right)$ are the matrices of the adjoint representation of the gauge group
\begin{equation}
g\,T_a\,g^{-1}=T_b\,d_{ba}\left(g\right)
\end{equation}

Due to the introduction of the extra scalar fields $h_{ab}$, the system described above has plenty of self-dual solutions. Using a spherically symmetric ansatz one can show in fact that the usual 't Hooft-Polyakov monopole \cite{thooft,polyakov} becomes a self-dual solution with a particular configuration of the $h$-fields. The system above is also conformally invariant in the three dimensional space $\relax\leavevmode\hbox{\rm I\kern-.18em R}^3$. Using an ansatz based on such conformal symmetry one construct  solutions with toroidal magnetic fields and vanishing magnetic charge. For more details on such results we refer to \cite{henrique}. 

\section{Skyrmions in $(3+1)$-dimensions}
\label{sec:skyrmions}
\setcounter{equation}{0}

Skyrmions are topological soliton solutions of theories in $(3+1)$-dimensions with target space being the group $SU(2)$. The three fields in $SU(2)$ are interpreted as the three pions $\pi^+$, $\pi^0$ and $\pi^-$. Such type solutions are interpreted, following a proposal of Skyrme \cite{skyrme1,skyrme2}, as nuclei and the topological charge plays the role of the baryonic number \cite{mantonbook,shnirbook}. 

The relevant topological charge in this case is given by the integral over the three dimensional space $\relax\leavevmode\hbox{\rm I\kern-.18em R}^3$
\begin{equation} 
Q_B= \frac{i}{48\,\pi^2}\,\int d^3x\; K\left(U\right)\,\varepsilon_{ijk}\,\widehat{\rm Tr}\left(R_i\,R_j\,R_k\right) 
\label{charge0}
\end{equation}
 with $R_i=i\, \partial_i U\,U^{\dagger}=R_i^a\,T_a$,  $U\in SU(2)$, and $K\left(U\right)$ is an arbitrary real functional of the chiral fields $U$, but not of their derivatives.  $K$ can be thought as a deformation of the metric on the target space $SU(2)$. We use the notation $\widehat{\rm Tr}\left(T_a\,T_b\right)=\delta_{ab}$, with $T_a$, $a=1,2,3$, being the generators of the Lie algebra of $SU(2)$.
 
 We now discuss some Skyrme type models with  exact self-dual sectors. In all cases the fields are real.  
 
 \subsection{The BPS Skyrme model} 
 
 Following (\ref{topcharge}) we introduce the real quantities
 \begin{equation}
 {\cal A}_{\alpha}\equiv \frac{\lambda}{24}\,\varepsilon_{ijk}\,\widehat{\rm Tr}\left(R_i\,R_j\,R_k\right) ;\qquad\qquad {\widetilde{\cal A}}_{\alpha}\equiv K=\mu\,\sqrt{V}
 \label{bpsskyrmesplit}
 \end{equation}
 where $\lambda$ and $\mu$ are coupling constants, and $V$ plays the role of the potential. Then one observes that (\ref{bpsskyrmesplit}) can be written in the form (\ref{topcharge}). The self-duality equations (\ref{sdeqs}) become 
 \begin{equation}
 \frac{\lambda}{24}\,\varepsilon_{ijk}\,\widehat{\rm Tr}\left(R_i\,R_j\,R_k\right)=\pm \mu\,\sqrt{V}
 \label{bpsskyrmesdeq}
 \end{equation}
 The energy functional (\ref{energy}) becomes
 \begin{equation}
 E=\int d^3x\, \left[\frac{\lambda^2}{24^2}\,B_i\,B_i+\mu^2\,  V\right]
\end{equation}
with $B_i=\varepsilon_{ijk}\,\widehat{\rm Tr}\left(R_i\,R_j\,R_k\right) $. Such a model was proposed in \cite{adam1} and has been applied in many contexts including nuclear physics and neutron stars \cite{adam2,adam_prl,adam_neutron_star}. The solutions of (\ref{bpsskyrmesdeq}) have been constructed using a spherically symmetric ansatz, for the potential $V={\rm Tr}\left(1-U\right)/2$, and they are of the compacton type, i.e. the fields go zero for a finite value of the radial distance. 

\subsection{A special self-dual Skyrme model}

Let us denote 
\begin{equation}
A_i=i\,\widehat{\rm Tr}\left(\partial_iU\,U^{\dagger}\,T_3\right);\qquad\qquad H_{ij}=\partial_iA_j-\partial_jA_i=i\,\widehat{\rm Tr}\left(\left\lbrack\,{\partial_iU\,U^{\dagger}}\, ,\,{\partial_jU\,U^{\dagger}}\,\right\rbrack\,T_3\right)
\label{ahdefshnir}
\end{equation}
with $U\in SU(2)$. Writing $R_i=i\, \partial_i U\,U^{\dagger}=R_i^a\,T_a$, we have that
\begin{equation}
\varepsilon_{ijk}A_i\,H_{jk}=2\,\varepsilon_{ijk}\,R_i^1\,R_j^2\,R_k^3=-i\,\frac{2}{3}\,\varepsilon_{ijk}\,\widehat{\rm Tr}\left(R_i\,R_j\,R_k\right)
\end{equation}
Taking $K=-4$, we can write (\ref{charge0}) as
\begin{equation}
Q_B=\frac{1}{4\,\pi^2}\,\int d^3x\, A_i\,B_i
\label{topchargeshnir}
\end{equation}
with
\begin{equation}
B_i=\frac{1}{2}\,\varepsilon_{ijk}\,H_{jk}
\end{equation}
We now introduce the real quantities
\begin{equation}
{\cal A}_{\alpha}\equiv m_0\, f\, A_i;\qquad\qquad\qquad {\widetilde{\cal A}}_{\alpha}\equiv \frac{1}{e_0\,f}\,B_i
\end{equation}
where $m_0$ and $e_0$ are coupling constants. Then we can write (\ref{topchargeshnir}), up to a constant, in the same form as (\ref{topcharge}). The self-duality equations (\ref{sdeqs}) become
\begin{equation}
m_0\, e_0\, f^2\, A_i=\pm B_i
\label{sdeqsshnir}
\end{equation}
and the energy functional (\ref{energy}) becomes
\begin{equation}
E=\frac{1}{2}\int d^3x\, \left[ m_0^2\,f^2\, A_i^2+\frac{1}{e_0^2\,f^2}\,B_i^2\right]
\label{energyshnir}
\end{equation}
Such a theory was first proposed in \cite{bpswojtek} for the case $f=1$, and then generalized in \cite{bpsshnir} for an arbitrary real function $f$. The theory with $f=1$ does not possess finite energy solutions in $\relax\leavevmode\hbox{\rm I\kern-.18em R}^3$ due to an argument by  Chandrasekhar \cite{chandra} in the context of force free fields in magnetohydrodynamics. However, exact solutions have been constructed in \cite{bpswojtek} for the case where the three dimensional space is the three sphere $S^3$. 

The self-duality equations (\ref{sdeqsshnir}) and the energy (\ref{energyshnir}) are conformally invariant in $\relax\leavevmode\hbox{\rm I\kern-.18em R}^3$, and this fact was used in \cite{bpsshnir} to build a conformal ansatz based on the toroidal coordinates
\begin{equation}
x^1= \frac{a}{p}\, \sqrt{z}\,\cos \varphi\;;\qquad\quad
x^2= \frac{a}{p}\, \sqrt{z}\,\sin \varphi\;;\qquad\quad
x^3= \frac{a}{p}\, \sqrt{1-z}\,\sin \xi
\label{toroidal}
\end{equation}
where
\begin{equation}
p=1-\sqrt{1-z}\,\cos\xi\qquad\qquad\qquad 0\leq z\leq 1\qquad\qquad 0\leq \varphi\, ,\,\xi\leq 2\,\pi
\label{pdef}
\end{equation}
Parameterizing the $SU(2)$ group elements as
\begin{eqnarray}
U=\left(
\begin{array}{cc}
Z_2& i\,Z_1\\
i\,Z_1^*& Z_2^*
\end{array}\right) \;;\qquad\qquad\qquad\qquad \mid Z_1\mid^2+\mid Z_2\mid^2=1
\end{eqnarray}
the vector $A_i$, introduced in (\ref{ahdefshnir}), can be written as 
\begin{equation}
A_{\mu}= \frac{i}{2}\left(Z_a^*\partial_{\mu} Z_a-Z_a\partial_{\mu} Z_a^*\right) 
\label{adef}
\end{equation}
The conformal ansatz corresponds to
\begin{equation}
Z_1= \sqrt{F(z)}\, e^{i\, n\,\varphi}\qquad\qquad\qquad
Z_2= \sqrt{1-F(z)}\, e^{i\,m\,\xi}
\label{parameterizez}
\end{equation}
with $m$ and $n$ being integers. The self-duality equations (\ref{sdeqsshnir}) are solved by the functions 
\begin{equation}
F=\frac{m^2\,z}{m^2\,z +n^2(1-z)} \qquad\qquad\qquad\qquad  f^2= \frac{2\,p}{m_0\,e_0\,a}~  \frac{\mid m\,n\mid}{[m^2\,z +n^2(1-z)]}
\label{solution-nm}
\end{equation}
The energy and the topological charge evaluated on such solutions are
\begin{equation}
E=4\,\pi^2\,\frac{m_0}{e_0}\mid m\,n\mid;\qquad\qquad \qquad Q=-m\,n
\end{equation}
We then have an infinite number of exact solutions, and they  correspond to special types of self-dual Skyrmions with target space $S^3\equiv SU(2)$.  Due to the conformal symmetry the solutions do not have a fixed size.  For more details we refer to \cite{bpsshnir}.

\subsection{A more general self-dual Skyrme model}

Using the fact that the quantites $R_i=i\, \partial_i U\,U^{\dagger}=R_i^a\,T_a$, satisfy the Maurer-Cartan equation
\begin{equation}
\partial_{\mu}R_{\nu}-\partial_{\nu}R_{\mu}+i\,\left\lbrack\,{R_{\mu}}\, ,\,{R_{\nu}}\,\right\rbrack=0
\label{maurercartan} 
\end{equation}
we can write the topological charge (\ref{charge0}), for $K=1$, as
\begin{eqnarray}
Q_B&=& \frac{i}{96\,\pi^2}\int d^3x\; \varepsilon_{ijk}\,\widehat{\rm Tr}\left(R_i\,\left\lbrack\,{R_j}\, ,\,{R_k}\,\right\rbrack\right)
 =-\frac{1}{96\,\pi^2}\int d^3x\; \varepsilon_{ijk}\,\widehat{\rm Tr}\left(R_i\,\left(\partial_j R_k-\partial_k R_j\right)\right) 
 \nonumber\\
 &=& -\frac{1}{48\,\pi^2}\int d^3x\; \varepsilon_{ijk}\,R^a_i\,\partial_j R^a_k
 \equiv -\frac{1}{48\,\pi^2}\,\frac{e_0}{m_0}\,\int d^3x\; {\cal A}_i^a\, {\widetilde{\cal A}}_i^a
 \label{topchargeleandro}
\end{eqnarray}
where we have introduced the real quantities
\begin{equation}
{\cal A}_i^a\equiv m_0\,R_i^b\, k_{ba} \; ; \qquad\qquad\qquad \qquad 
{\widetilde{\cal A}}_i^a\equiv \frac{1}{e_0}\,k^{-1}_{ab}\,\varepsilon_{ijk}\,\partial_j R^b_k
\end{equation}
where $k_{ab}$ is some invertible matrix, and $m_0$ and $e_0$ are coupling constants. Therefore, the topological charge (\ref{charge0}), for $K=1$, can be written in the same form as (\ref{topcharge}). The self-duality equations (\ref{sdeqs}) become
\begin{equation}
\lambda \,h_{ab}\,R^b_i=\frac{1}{2}\,\varepsilon_{ijk}\,H^a_{jk}\qquad\qquad {\rm with} \qquad \qquad\lambda=\pm\,m_0\,e_0
\label{selfdualleandro}
\end{equation} 
where $h=k\,k^T$ is a real, symmetric and invertible matrix, and where we have denoted 
\begin{equation}
H^a_{ij}=\partial_iR_j^a-\partial_jR_i^a=\varepsilon_{abc}\,R_{\mu}^b\,R_{\nu}^c
\end{equation}
The energy functional (\ref{energy}) becomes
\begin{equation}
E= \int d^3x\left[ \frac{m_0^2}{2}\, h_{ab}\,R^a_{i}\,R^{b}_{i}+\frac{1}{4\,e_0^2}\, h^{-1}_{ab}\,H^a_{ij}\,H^{b}_{ij}\right]
\label{e1}
\end{equation}
The energy evaluated on the self-dual solutions of (\ref{selfdualleandro}) is given by
\begin{equation} 
E=48\,\pi^2\,\frac{m_0}{e_0}\,\mid Q\mid 
\label{e1top}
\end{equation} 
Such a theory was proposed in \cite{laf2017} and further explored in \cite{us}. The entries of the matrix $h_{ab}$ are considered as six real scalar fields added to the theory. Note that for $h=\hbox{{1}\kern-.25em\hbox{l}}$ the model reduces to the original Skyrme model \cite{skyrme1,skyrme2}. The topological charge (\ref{topchargeleandro}) is interpreted, following Skyrme, as the baryon number. More recently such a model was extended by treating a fractional power of the density of the topological charge as an order parameter to describe a fluid of baryonic matter \cite{nuclear}. Such an extension has lead to a very interesting application to nuclear theory. The model describes with quite good accuracy the binding energies per nucleon of more than 240 nuclei, and also the relation between their radii and baryon number. 

The important results of \cite{us}   are: {\em i)} the first order self-duality equations (\ref{selfdualleandro}) imply the nine static second order Euler-Lagrange equations associated  to  fields $U$ and $h_{ab}$, {\em ii)}  the static Euler-Lagrange equations associated to the fields $h_{ab}$ are equivalent to the self-duality equations, {\em iii)}  given a configuration for the $U$-fields one can solve the self-duality equations by taking $h_{ab}$ to be
\begin{equation}
h=\frac{\sqrt{{\rm det}\,\tau}}{ m_0\,e_0}\; \tau^{-1};\qquad\qquad {\rm with}\qquad \qquad\tau_{ab}= R_{i}^a\,R_{i}^b
\end{equation} 
So, the fields $h_{ab}$ are spectators in the sense that they adjust themselves to solve the self-duality equations for any configuration of the $U$-fields. Note that the matrix $\tau$ is similar to the Skyrme model strain tensor  \cite{mantonbook}. For  $U$-field configurations where $\tau$ is singular the matrix $h_{ab}$ still solves the self-duality equation but it is not completely determined by $U$, and have some arbitrary components \cite{us}. The theory (\ref{e1}) is conformally invariant in the three dimensional space $\relax\leavevmode\hbox{\rm I\kern-.18em R}^3$ and that plays an important role in the properties of the model. Exact solutions to the self-duality equations (\ref{selfdualleandro}) have been constructed in \cite{us} using an holomorphic ansatz, and also a toroidal ansatz based on the conformal symmetry. For more details about these results we refer to \cite{laf2017,us,nuclear}. 

\section{Instantons in four Euclidean dimensions}
\label{sec:instantons}
\setcounter{equation}{0}

As a last example of  applications of the methods described in Section \ref{sec:introduction} we just mention the case of instanton solutions of Yang-Mills theory in four Euclidean dimensions. The relevant topological charge in this case is the Pontryagin number
\begin{equation}
Q_{YM}=\int d^4x\, {\rm Tr}\left(F_{\mu\nu}\,{\widetilde F}^{\mu\nu}\right)
\label{topchargeym}
\end{equation}
with $F_{\mu\nu}$ being the filed tensor and ${\widetilde F}_{\mu\nu}$ its Hodge dual, i.e.
\begin{equation}
F_{\mu\nu}=\partial_{\mu}A_{\nu}-\partial_{\nu}A_{\mu}+i\,e\,\left\lbrack\,{A_{\mu}}\, ,\,{A_{\nu}}\,\right\rbrack;\qquad\qquad \qquad
{\widetilde F}_{\mu\nu}=\frac{1}{2}\,\varepsilon_{\mu\nu\rho\sigma}\,F^{\rho\sigma}
\end{equation}
and $A_{\mu}$ being the gauge potential for a compact Lie group $G$. Following (\ref{topcharge}) we denote
\begin{equation}
{\cal A}_{\alpha}\equiv F_{\mu\nu};\qquad\qquad \qquad{\widetilde{\cal A}}_{\alpha}\equiv {\widetilde F}_{\mu\nu}
\end{equation}
The self-duality equations (\ref{sdeqs}) become
\begin{equation}
F_{\mu\nu}=\pm {\widetilde F}_{\mu\nu}
\label{sdeqsym}
\end{equation}
and the functional (\ref{energy}) becomes the Yang-Mills Euclidean action
\begin{equation}
S_{YM}=\frac{1}{8}\int d^4x\,\left[{\rm Tr}\left(F_{\mu\nu}F_{\mu\nu}\right)+{\rm Tr}\left({\widetilde F}_{\mu\nu}{\widetilde F}_{\mu\nu}\right)\right]=
\frac{1}{4}\int d^4x\,{\rm Tr}\left(F_{\mu\nu}F_{\mu\nu}\right)
\end{equation}
where we have used the fact that ${\rm Tr}\left(F_{\mu\nu}F_{\mu\nu}\right)={\rm Tr}\left({\widetilde F}_{\mu\nu}{\widetilde F}_{\mu\nu}\right)$. 

The solutions of (\ref{sdeqsym}) are the well known instanton solution of Euclidean Yang-Mills theory, and they plays an important role in the structure of the vacua and also on non-perturbative phenomena in Yang-Mills theory \cite{belavininstanton,mantonbook}. 

Following (\ref{newsplitting}) one could introduce a real, symmetric and invertible matrix $h_{ab}$ into the self-duality equations (\ref{sdeqsym}) as $F_{\mu\nu}^b\, h_{ba}=\pm {\widetilde F}_{\mu\nu}^a$, with $F_{\mu\nu}=F_{\mu\nu}^a\, T_a$, and ${\widetilde F}_{\mu\nu}={\widetilde F}_{\mu\nu}^a\,T_a$, and $T_a$, $a=1,2,\ldots {\rm dim}\, G$, being a basis for the Lie algebra of the gauge group $G$. However, due to arguments similar to those used  (\ref{sdeqscpn2})-(\ref{honecpn}), one can show that such a matrix $h$ has to be the unity matrix \cite{henriquesdym}.

\vspace{2cm}

{\bf Acknowledgements:} The author is supported by Funda\c c\~ao de Amparo \`a Pesquisa do Estado de S\~ao Paulo (FAPESP) (contract 2022/00808-7), and Conselho Nacional de Desenvolvimento Cient\'ifico e Tecnol\'ogico - CNPq (contract 307833/2022-4).

\end{document}